\documentclass[conference]{IEEEtran}

\usepackage{cite}
\usepackage{amsmath,amssymb,amsfonts}
\usepackage{algorithmic}
\usepackage{graphicx}
\usepackage{textcomp}
\usepackage{xcolor}
\usepackage{subfigure}
\usepackage{url}
\usepackage[normalem]{ulem}
\def\BibTeX{{\rm B\kern-.05em{\sc i\kern-.025em b}\kern-.08em
    T\kern-.1667em\lower.7ex\hbox{E}\kern-.125emX}}
    
\usepackage[absolute]{textpos}

\begin{document}

\begin{textblock}{13}(1.5,0.2)
{\footnotesize \noindent S. Pulipati, V. Ariyarathna, U. D. Silva, N. Akram, E. Alwan, A. Madanayake, S. Mandal, and T. S. Rappaport, ``A Direct- Conversion Digital Beamforming Array Receiver with 800 MHz Channel Bandwidth at 28 GHz using Xilinx RF SoC," \textit{2019 IEEE International Conference on Microwaves, Communication, Antennas  and Electronic Systems (COMCAS)}, Tel Aviv, Israel, Nov 2019, pp. 1-5.}
\end{textblock}

\title{A Direct-Conversion Digital Beamforming Array Receiver with 800 MHz Channel Bandwidth at 28 GHz using Xilinx RF SoC\\
}

\author{\IEEEauthorblockN{Sravan Pulipati, Viduneth Ariyarathna,\\ Udara De Silva, Najath Akram, \\Elias Alwan, Arjuna Madanayake}
\IEEEauthorblockA{\textit{Dept. of Elec. and Comp. Engineering} \\
\textit{Florida International University}\\
Miami, FL, USA \\ 
amadanay@fiu.edu}
\and
\IEEEauthorblockN{Soumyajit Mandal}
\IEEEauthorblockA{\textit{Dept. of Elec. Eng. and Comp. Science} \\
\textit{Case Western Reserve University}\\
Cleveland, OH, USA \\
sxm833@case.edu  }
\and
\IEEEauthorblockN{Theodore S. Rappaport}
\IEEEauthorblockA{\textit{Dept. of Elec. and Comp. Engineering} \\
\textit{New York University}\\
New York,NY, USA \\
tsr@nyu.edu}
}

\maketitle

\begin{abstract}
This paper discusses early results associated with a fully-digital direct-conversion array receiver at 28~GHz. The proposed receiver
makes use of commercial off-the-shelf (COTS) electronics, including the receiver chain. The design consists of a custom 28~GHz patch antenna sub-array providing gain in the elevation plane, with azimuthal plane beamforming provided by real-time digital signal
processing (DSP) algorithms running on a  Xilinx Radio Frequency System on Chip (RF SoC). The proposed array receiver employs
element-wise fully-digital array processing that supports ADC sample rates up to 2~GS/second and up to 1~GHz of operating  bandwidth per antenna. The RF mixed-signal data conversion circuits and DSP algorithms operate on a single-chip RF SoC solution installed on the Xilinx ZCU1275 prototyping platform.

\end{abstract}


\section{Introduction}
It is an exciting time for wireless communications. The rush to move to 5G in the 24/28 GHz bands (and eventually the 39 GHz band as well) has resulted
in fierce competition among network providers and original equipment manufacturers (OEM) to provide highly-adaptable, high-performance hardware and circuit solutions that take full advantage of the mm-wave spectrum. There are massive opportunities for new applications in the mm-wave space waiting to be exploited, and new networks having high throughput, capacity and low latency are the starting point for the next wave of `Ubers, Facebooks, and Instagrams of the future'~\cite{ted_5g,ted_6g}.
 The networks of the future, at 28 GHz and beyond, will be based on electronic systems that heavily depend on the integration of radio-frequency (RF) electronics with digital back-ends that run
  software-defined radio (SDR) algorithms. Proposed sixth-generation (6G) networks will likely include a significant amount of artificial intelligence (AI) and RF machine learning (RF-MLS) algorithms.
  Regardless of the software approach, whether linear time invariant (LTI) or non-linear/ML/AI based, the ultimate performance of the network is still grounded on the performance of the analog front-ends that interface the
  information-carrying signals to the physical world. Components such as antennas, low-noise amplifiers (LNAs), and power amplifiers (PAs) for example, will continue to play an important role in emerging 5G/6G systems.
  The design of the mm-wave RF front-end is critically important because any loss of front-end performance is difficult, if not impossible, to correct in the digital post-processing back-end. 

\begin{figure}[!t]
\centering
\scalebox{0.9}{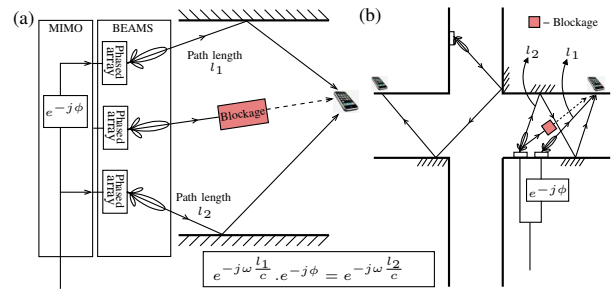}
\vspace{-2ex}
\caption{Possible applications of MIMO + multi-beam beamforming (a) in a blocked LOS path; (b) to improve SNR when LOS is available by using multi-path techniques, and to connect across corners in an urban environment. \label{ray_d}}
\vspace{-4 mm}
\end{figure}

Next-generation (6G) wireless access points will employ a combination of multi-input-multi-output (MIMO) theory with multi-beam fully-digital beamforming. 
The high-gain array factors provided by very sharp beams will be required for overcoming high path losses and also mitigating environment attenuation due to absorbing gases (e.g., oxygen at 60 GHz), rain, hail, dust, and other opaque objects (trees, humans) that degrade signal-to-noise ratio (SNR) at the receiver. Future networks will thus exploit both the massive bandwidth available in the mm-wave bands, and also the high SNR available due to the use of high-gain antenna arrays to 
achieve spectral efficiency via orthogonal frequency division multiplexing (OFDM). In particular, the high SNR that can be achieved with digital beamforming will allow quadrature amplitude modulation (QAM) with up to 
1024 discrete constellation points per OFDM sub-carrier. In addition, the combination of MIMO techniques with fully-digital multi-beam beamforming allows coherent combination of several ray-like channels (from the sharp beams) to achieve several important capabilities, such as 1) connecting when line-of-sight (LOS) is not available due to channel blockage, 2) connecting across corners in a densely built environment (e.g., downtown Miami/NYC), and 3) coherent combination of multiple ray-like channels to improve SNR even when LOS is available. These scenarios are illustrated in Fig.~\ref{ray_d}~\cite{iwat2019}.

\begin{figure}[!ht]
\centering
\scalebox{1.1}{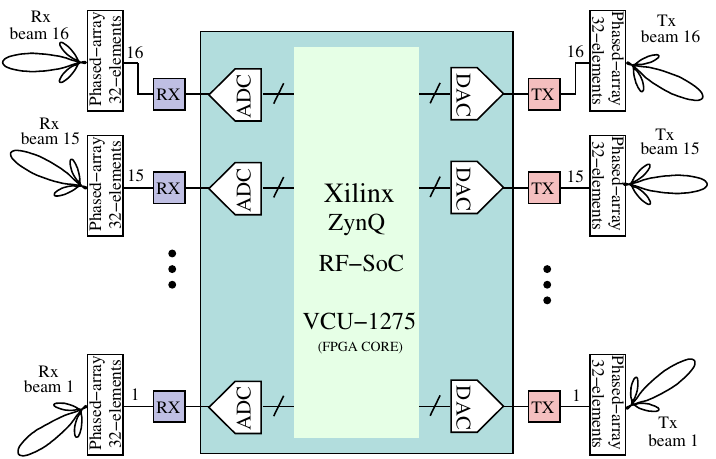}
\vskip -1ex
\caption{System overview: a 16$\times$16 beamforming+MIMO access point for 28 GHz OFDM. \label{ov_fig}}
\vskip -1ex
\end{figure}

In this paper, we discuss the implementation of  a four-element uniform linear array (ULA) in receive mode. Each antenna element of this array operates at a center frequency of 28 GHz and provides a bandwidth of about 800 MHz. To the authors' knowledge, this is the first practically realized and experimentally verified antenna array in this frequency range that provides such high bandwidth.
The paper provides full design schematics of the 28 GHz receiver, example antenna designs, and details of the Xilinx RF SoC-based DSP back-end used for real-time signal processing at the physical layer.


\section{Design Specifications and Constraints}
This section discusses the technical specifications of the receiver, which is designed to support 64-QAM modulation for 5G OFDM-based wireless communications. A description of the design procedure can be found in~\cite{RF_HFIC}.

The proposed receiver design is designed to operate in the 27.5 GHz-28.35 GHz frequency range, which has been allocated for commercial 5G applications in the United States~\cite{fcc_5g_28}. The OFDM parameters listed below were used for the theoretical calculations as well as to model the system using AWR Microwave Office mm-wave (mmW) circuits and systems simulator software: FFT size ($N_{FFT}$): 512, modulation ($M$): 64 QAM, guard interval (GI): $1/8$, data subcarriers ($N_{D}$): 336, and subcarrier spacing (CS): 1.65 MHz.
These system parameters provide the required bandwidth (800 MHz) and a maximum data rate of
	\begin{align}\label{d_rate}
		R_b = \frac{\log_2(M) \times N_D \times \text{CS} }{1+ \text{GI}} &= \frac{6 \times 336 \times 1.65 \times 10^6}{1+0.125} \nonumber \\
		& \simeq 3 ~\text{Gbps}. 
	\end{align} 

To establish a wireless communication link with bit error rate (BER) better than $10^{-5}$ for $M=64$, $E_b/N_0$ should be at least 17.8 dB~\cite{sklar1988digital}. Hence, the required SNR at the demodulator inputs is given by
\begin{align}\label{snrreq}
SNR_{dB} &= \left( \frac{E_b}{N_0}\right) +10\log_{10}\left(\frac{\log_2(M) N_D}{(1+ \text{GI})N_{FFT}}\right) = {23.2~ dB}. \nonumber 
\end{align}
The receiver chain is designed by taking this SNR value as a design constraint. 

\section{Front-End Design}
A system overview of the mmW front-end design is shown in Fig.~\ref{ov_fig}~\cite{iwat2019}. Note that our current implementation only uses 4 elements. However, we plan to scale the system to 32 elements in the near future.

\subsection{Cascaded Noise Figure and Gain}
The performance of the mm-wave  receiver is optimized to have a low noise figure (NF) and high gain. The first LNA, which is directly connected to the antenna, largely sets the overall sensitivity and NF of the receiver. Thus it should have the lowest NF in the chain. The best trade-off between the NF and gain of the LNA can be calculated using well-known microwave receiver design methods available in the literature~\cite{mcclaning,pozar_rf}.
A complete analysis of the maximum possible noise figure $F_{cas}$ and minimum possible gain $G_{cas}$ for the receiver chain to satisfy SNR$=23.2$~dB is described in~\cite{iwat2019}. The calculations were made under the assumptions that 1) the signal levels at the input of the antenna are in the range of $-85$~dBm, and 2) each antenna of a 32 element array provides an individual gain of 15 dB. The reported noise figure and gain value requirements are 5.8 dB and 55 dB respectively. 

%
%
%
%
%

%
\begin{figure}[t]
\centering
\includegraphics[scale = 0.4]{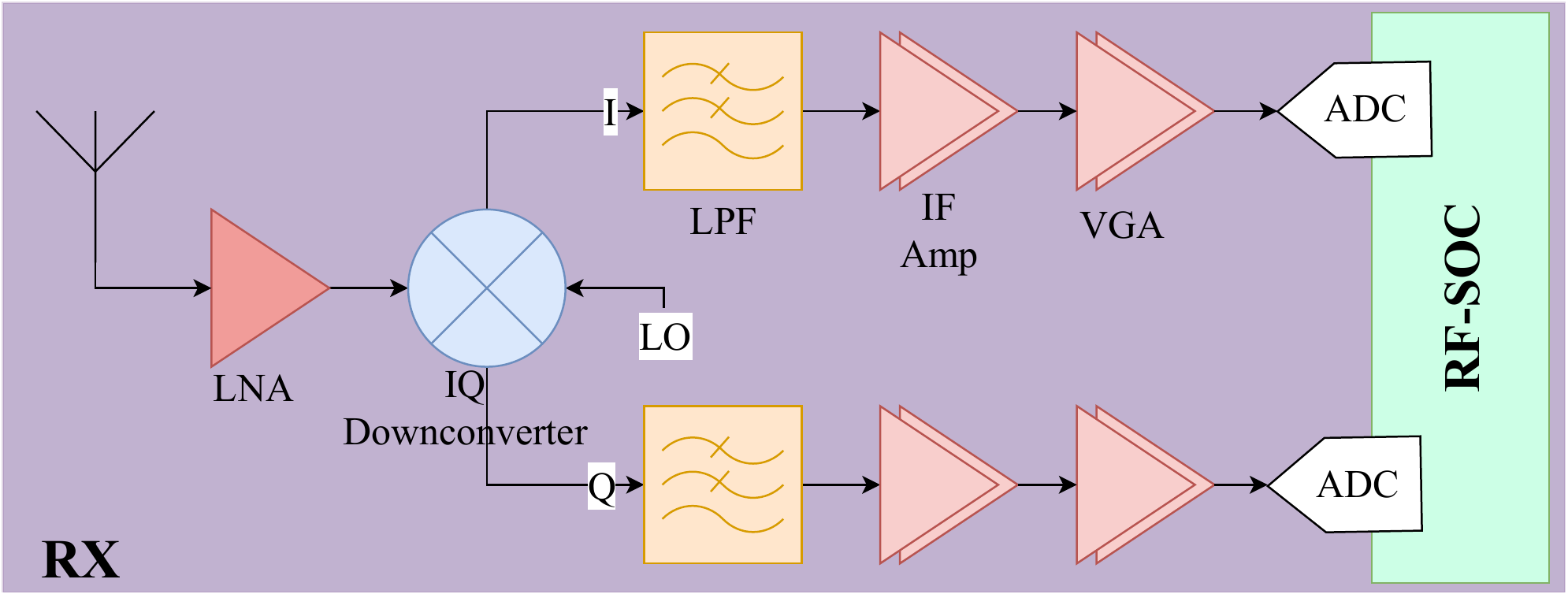}
\vspace{-2ex}
\caption{Architecture of the proposed mmW receiver.\label{ov_fig_2}}
\end{figure}
%
%
%

\subsection{Architecture and Component Selection}
The proposed mmW receiver chain employs a superheterodyne IQ receiver architecture as shown in Fig.~\ref{ov_fig_2}. The narrow-band nature of patch antennas eliminated the need for separate band-pass filters. A low-side LO is used at the IQ-downconverter (HMC1065LP4E) to generate an intermediate frequency (IF) signal of 0.15-1 GHz, which is further amplified and sampled by high speed RF-SoC ADCs. Careful selection of components is critical for providing the desired NF and gain values. Components with higher third-order intercept (TOI) levels are desirable to provide better linearity and thus higher dynamic range (DR). The final component selections are shown in Table~\ref{Tab_comp2}.

\begin{table}[h!] 
\renewcommand{\arraystretch}{1.3}
\vspace{-4ex}
\caption{Components and specifications for the mmW receiver}
\label{Tab_comp2}
\centering 
\begin{tabular}{|c|c|c|c|}
\hline
\textbf{Component} & \textbf{Gain} & \textbf{NF} & \textbf{OIP3} \\ 
 & (dB) & (dB) & (dBm) \\ \hline
LNA ({MAAL-01111})        & 19                 & 2.5              & 20           \\ \hline

Down converter ({HMC1065LP4E})        &  9                 & 3               & 14                 \\ \hline
Low-pass filter (LPF) ({LFCN-900+})          & $-1$                 & 1                &        N/A             \\ \hline
IF amplifier ({RAM-8A+})            & [31.5,24]                 & 2.6              & 24.4               \\ \hline
VGA ({ADL5331})            & [$-15$,15]                 & 9              &  39               \\ \hline
\end{tabular}
\end{table}

Ideal values of cascaded gain and NF using the selected components are computed using standard formulas~\cite{mcclaning}, resulting in $G_{cas,dB}=70$~ dB and $F_{cas,dB}=2.5$~dB. For the gain calculation, we considered 1) the maximum gain of the variable gain amplifier (VGA), and 2) the mid-band gain of the IF amplifier . The receiver's noise performance is well within the NF requirements, and with the help of VGA (ADL5331) it also provides 30~dB of adjustable gain to increase DR, i.e., keep the amplified signal levels within the full-scale range of the ADCs. Next, we evaluate and verify the performance of the proposed mmW receiver.

\section{Performance Evaluation}
To evaluate the real-world performance of the selected components, a simulation test-bench was set up in AWR Microwave Office simulation software.  A 64 QAM OFDM modulation scheme is employed, and the system is excited using only one OFDM subcarrier as the others were undergoing phase rotation. Correcting for this effect requires careful calibration, which will be undertaken in future work.
Since only one OFDM subcarrier is considered, applying $N_D=1$ in~\eqref{d_rate} and recomputing, the detectable signal $P_{in}$ at the antenna would be lowered to $-110$~dBm.

\begin{figure}[!ht]
\centering
\includegraphics[scale = 0.5]{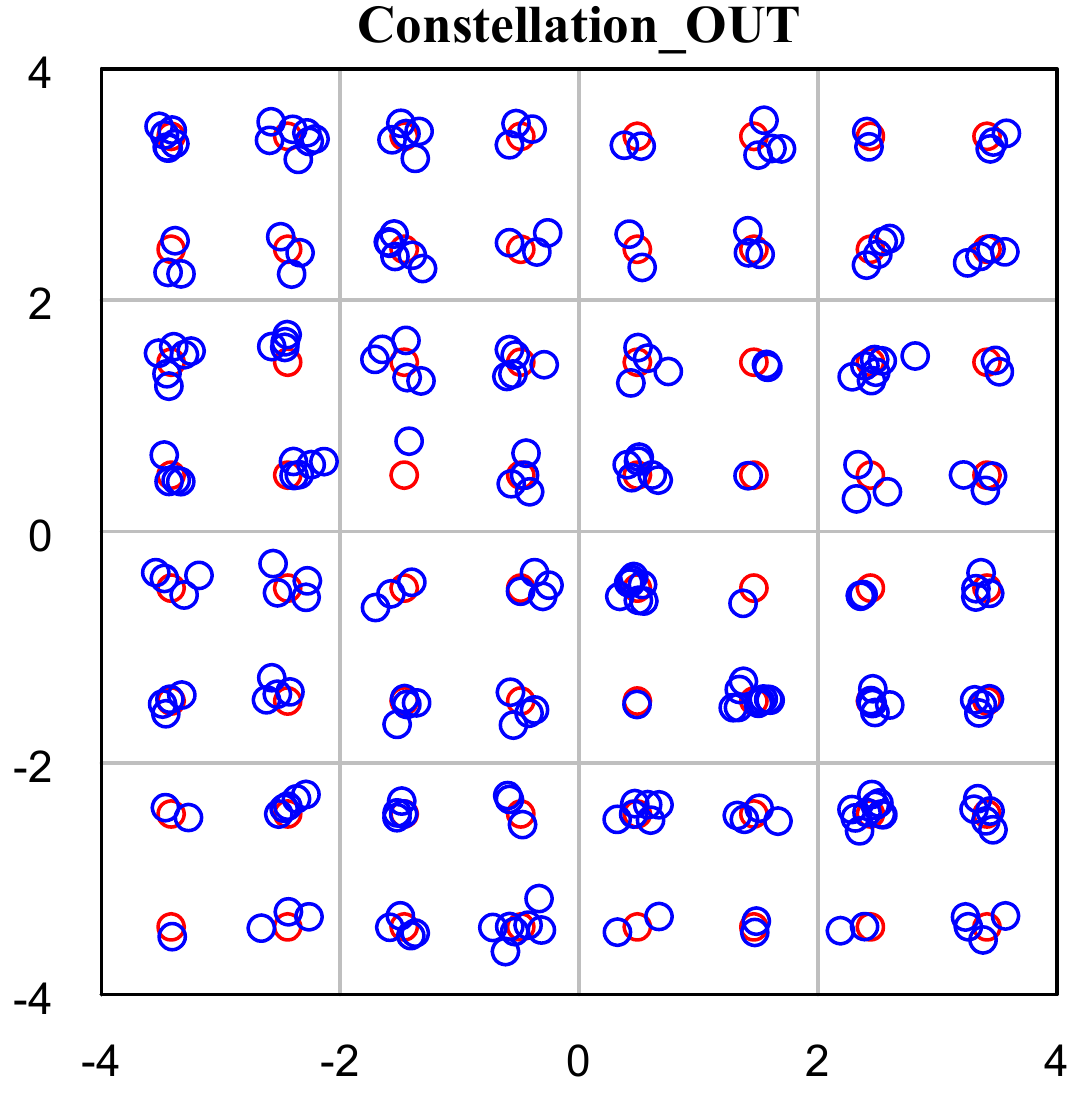}
\vskip -2ex
\caption{Simulated constellation for a 64 QAM single sub-carrier at the receiver side for an input level of $-110$~dBm. Red and blue circles indicate the ideal and measured values, respectively. \label{constellation}}
\end{figure}

The AWR model is simulated with a receiver power of $-110$~dBm at its input. At the center frequency of the IF band ($f_{Rf}-f_{LO}=0.6$~GHz), the values of $F_{cas}, G_{cas} \text{ and } OIP3_{cas}$ were found to be 4.7~dB, 49.4~dB, and 35.6~dBm respectively. Fig.~\ref{constellation} compares the constellation simulated after the QAM demodulator with an ideal 64-QAM constellation. No appreciable difference was observed between the ideal and simulated constellations. A numerical assessment is being conducted to quantify how closely the behavior of the real-world components matches those used in an ideal receiver.

\section{28 GHz Receiver Array Setup}
We use a custom 8-element series fed sub-array that employs a series-feeding structure to decrease the width of the field patterns in the vertical plane (elevation) so as to suppress interference and spurious signals, thus providing more directive gain. Additionally, designing a series-fed array with tapering of the patch widths helps in significantly reducing the side lobe levels (SLLs) of the field patterns along the axis of the array. In series-feeding, elements are spaced a guided wavelength apart along a uniform transmission line. By tapering the widths from
one edge to center element in ascending order, excitation is maximum at the center and decreases as one approaches the edge, which results in a slight increase in beam-width and reduction of gain when
compared to uniform excitation. The 28 GHz 8-element array is designed with a linear tapering having $-6$ dB pedestal (to control the SLL). The patch array is analyzed using a transmission line model~\cite{balanis}  in conjunction with wavelength-apart series-fed analysis~\cite{sainati}, to compute the dimensions of each patch that produce the corresponding amplitude excitation.

\begin{figure*}[h!]
\centering
\scalebox{0.7}{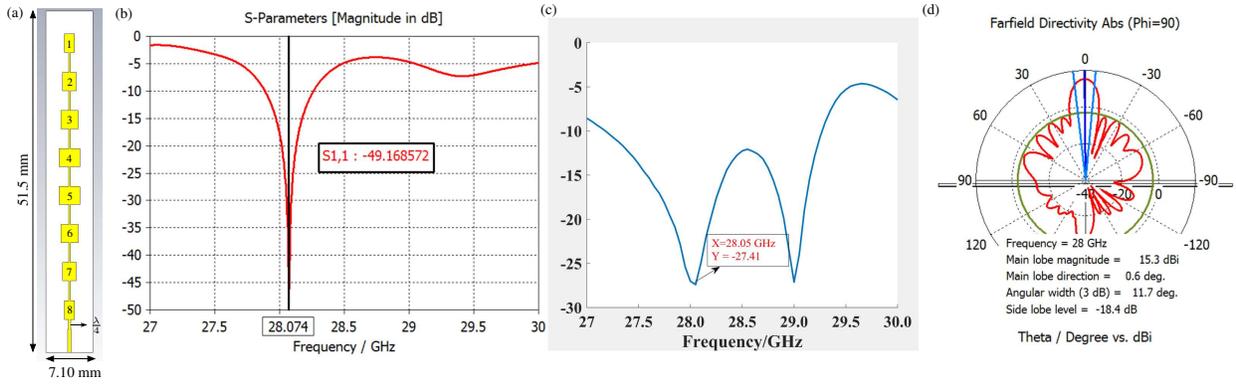}
\vskip -2ex
\caption{ (a) Dimensions of the antenna sub-array; (b) CST-simulated return loss ($S_{11}$); (c) Measured return loss ($S_{11}$); and (d) CST-simulated polar pattern of the sub-array at 28 GHz. Note that the polar pattern is along the axis of the array i.e., in the elevation plane.  \label{Fig:Simulations}}
\end{figure*}

Following the analysis described above, design dimensions for each patch were calculated for the board specifications shown in Table~\ref{Table_specs}. The materials were chosen to optimize performance at mm-wave frequencies, and also to account for practical element and transmission line dimensions at
these frequencies. The sub-array is designed in CST antenna simulation software. The optimized dimensions for each patch are shown in Table~\ref{patch_dimensions}, and its geometry is shown in Fig.~\ref{Fig:Simulations} (a). Matching to the 50~$\Omega$ feed is based on a quarter-wave
transformer. Simulated and measured results for $|S_{11}|$ are shown in Fig.~\ref{Fig:Simulations}~(b) and (c) respectively, whereas far-field patterns (vertical plane) are shown in Fig.~\ref{Fig:Simulations}(d). The proposed antenna resonates at 28.05 GHz with a return loss of 27.41 dB. The tapered 28 GHz array results in a SLL less than $-18$~dB, compared to $-13$ dB for uniform excitation. 

\begin{table}[]
\centering
\caption{Specifications for the Patch Antenna Design \label{Table_specs}}
\vskip -1ex
\begin{tabular}{|c|c|}
\hline
Frequency $(f_0)$                  & 28 GHz               \\ \hline
Substrate                          & RO4350B             \\ \hline
Dielectric constant ($\epsilon _r$) & 3.66                 \\ \hline
Dielectric height (h)             & 0.254 mm             \\ \hline
\end{tabular}
\vskip -2ex
\end{table}

\begin{table}[]
\centering
\caption{Dimensions for the Patch Antenna Design \label{patch_dimensions}}
\vskip -1ex
\begin{tabular}{|c|c|c|c|c|}
\hline
\textbf{Patches} &  \textbf{1st, 8th} & \textbf{2nd, 7th} & \textbf{3rd, 6th} & \textbf{4th, 5th} \\
\hline
Width & 1.55 & 2.06 & 2.59 & 3.12 \\ 
\hline
Length & 2.81 & 2.77 & 2.74 & 2.72  \\ \hline
\end{tabular}
\vskip -2ex
\end{table}

\begin{figure}[h!]
\centering
\includegraphics[scale = 0.60]{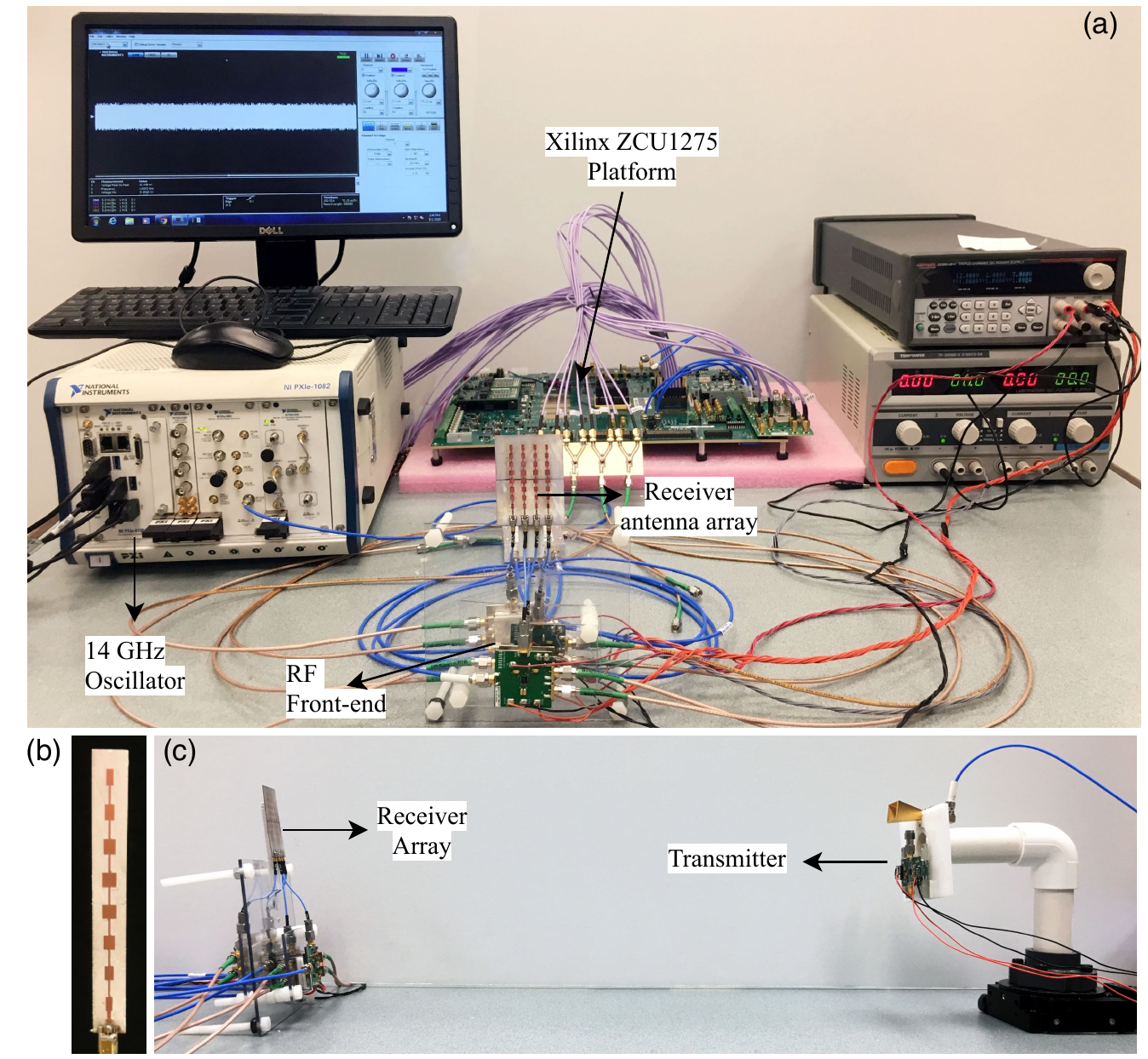}
\caption{ (a) The receiver setup for measurements; (b) Close-up of the fabricated antenna sub-array; and (c) The transmit-receive link.  \label{Fig:setup}}
\end{figure}

\subsection{Antenna Array Design}
Design of the 8-element series fed sub-array on CST was followed by the realization with an RF front end. The entire RF receiver setup is shown in Fig.~\ref{Fig:setup}(a) and a close up of the fabricated antenna is shown in Fig.~\ref{Fig:setup}(b).
The 4-element ULA is built using four sub arrays with a separation of 0.75 wavelengths (8.0~mm at 28~GHz). Each sub array is fed with a separate RF front-end as described in Fig.~\ref{ov_fig_2} using the components selected in Table~\ref{Tab_comp2}. A central LO is divided using a four-way splitter and used as the LO for the individual mixers. The single-ended outputs of each front-end are converted to differential pairs in order to drive the Xilinx RF SoC's built-in ADCs. The conversion uses RF baluns built using Mini-circuits TC1-1-13M+ RF transformers~\cite{baluns}, which have 1~dB insertion loss for signals from 4.5-1000 MHz.

\begin{figure*}[t!]
\centering
\vskip - 1ex
\scalebox{1}{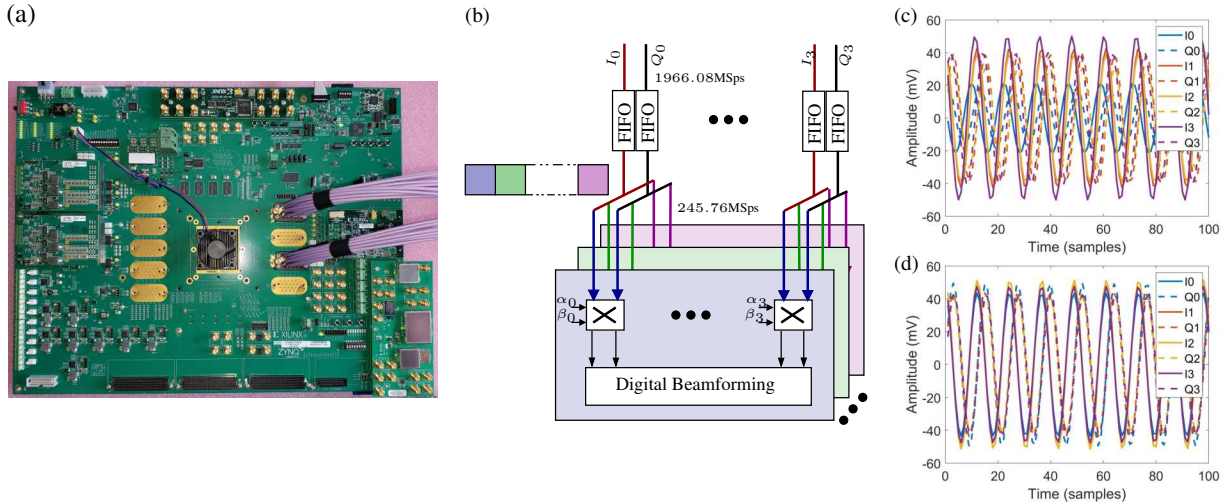}
\vskip -2ex
\caption{(a) The Xilinx ZCU1275 platform, which allows use of the XCZU29DR-2FFVF1760E Zynq UltraScale+ RFSoC chip; (b) Architecture of the digital back-end; (c) The sampled I and Q channels of each baseband signal; and (d) The digitally-calibrated channels fed invised to the digital beamforming cores.}
\vskip -2ex
\label{Fig:digitalbackend}
\end{figure*}

\subsection{Digital Back-End}
A Xilinx ZCU1275 evaluation board is used as the digital processing back-end of the 28-GHz array receiver. The board contains a XCZU29DR-2FFVF1760E Zynq UltraScale+ RFSoC chip, which integrates high-speed data converters along with a programmable logic fabric and application processing unit. Specifically, the chip supports 16 ADC channels upto 2 GSPS and 16
DAC channels upto 4 GSPS. Fig.~\ref{Fig:digitalbackend}(a) shows a close up of the Xilinx ZCU 1275 board, which provides Bulls Eye interfaces to access all the RFSoC's data converters. A Xilinx Analog Super Clock Module (HW-CLK-102)~\cite{1275UserGuide} that supports the ADC/DAC clock requirements is used to clock the data converters. The 16 ADC inputs in the RFSoC are arranged into 4 tiles, each of which samples 4 channels. The HW-CLK-102 module supports four differential RF clocks for the ADCs and three phase-aligned reference clocks for synchronization.

The data converters were configured to sample at 1966.08~MHz. To handle the total bandwidth of 800~MHz, the digital circuits were designed in a polyphase architecture with 8 parallel beamforming cores.  Therefore, the built-in FIFOs of the Xilinx Data Converter IP (XDCIP) core were configured to output a sample rate of $1966.08/8 = 245.76$~MSps with 8 sampled words per clock edge streamed into the beamforming cores. An overview of the digital back-end architecture is shown in Fig.~\ref{Fig:digitalbackend}(b). The outputs of each FIFO stream were synchronized to a single reference clock of 245.76~MHz that was derived from the analog sampling clock. In this way, the design can use 8 parallel digital cores to process the entire sampled bandwidth, as shown in Fig.~\ref{Fig:digitalbackend}(b). 

\subsection{Calibration}
The calibration mode of each ADC channel was set to ``Mode-2''~\cite{DataConverterIP_doc} in the XDCIP. In this mode, ADC calibration is handled by the start-up finite-state-machine of the XDCIP. In addition, the RF front-ends were digitally calibrated using gain and phase correction. For this purpose, we added a complex multiplier at each phase of each channel as shown in Fig.~\ref{Fig:digitalbackend}(b). The gain and phase mismatches with respect to a reference receiver were pre-measured using a reference signal. Complex calibration constants $\alpha_i + j\beta_i$, where $i=\{0,1,2,3\}$ and $\{\alpha, \beta \} \in \mathbb{R}$, were then estimated for each channel. Typical I and Q baseband channel data measured for a reference input signal before calibration are shown in Fig.~\ref{Fig:digitalbackend}(c). Fig.~\ref{Fig:digitalbackend}(c) shows the digitally-calibrated versions that are fed into the digital beamforming cores.

\subsection{Real-Time Beamforming and Measurement Setup}
\begin{figure}[t!]
\begin{center}
\includegraphics[scale=0.07]{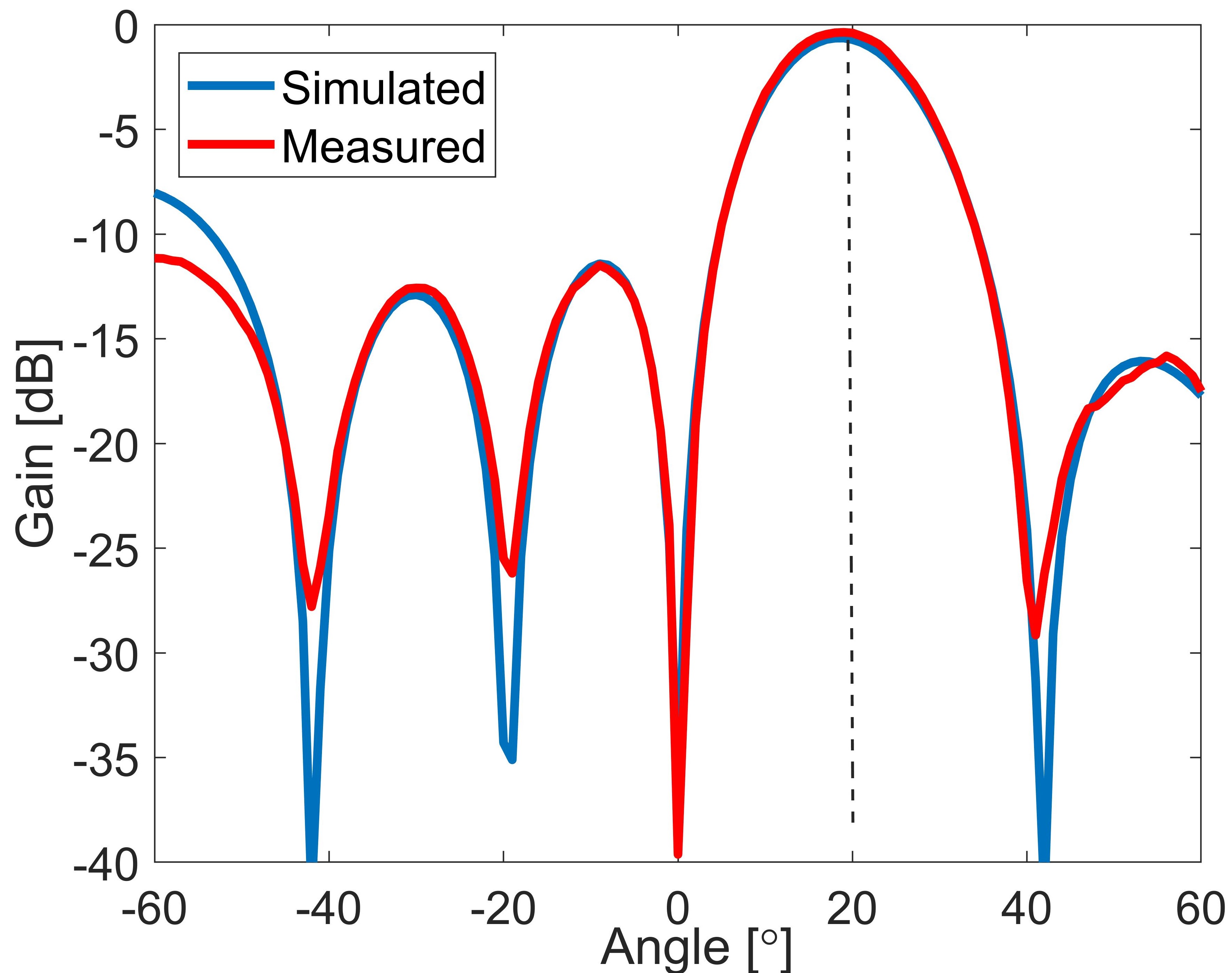}
\end{center}
\vskip -3 ex
\caption{Simulated versus measured array factors for an IF frequency of 300~MHz.}
\label{Fig_sim_meas_beams}
\vskip -1ex
\end{figure}

Fig.~\ref{Fig:setup}(a) shows the entire 28-GHz transmitter and receiver array setup used for measuring beams in real-time beams. As shown in Fig.~\ref{Fig:setup}(c), a transmit horn antenna is used to send out a 28.2 GHz carrier. The LO frequency is set to 27.9 GHz, thus producing 300 MHz IF signals. As a proof of concept, a simple weight-and-sum beamforming algorithm was implemented digitally on each of the 8 phases. The beamformer weights are set to pass signals arriving at $20^{\circ}$. The beamformed outputs from each phase are then used to compute the received energy for each direction of arrival. The angle of arrival of the 28-GHz wave front is changed by rotating the receiver array to obtain the array factor of the desired beam.  Fig.~\ref{Fig_sim_meas_beams} compares the measured and simulated array factors; the simulated plot accounts for the azimuthal element pattern of each patch antenna sub-array. 
The two plots are well-matched with respect to beam direction as well as side lobe performance, thus validating the proposed mmW array receiver.


\section{Conclusion}
We have described a fully-digital direct conversion 4-element array receiver operating at 28 GHz that uses a MIMO multibeam beamforming architecture for high- throughput 5G applications. Complete system analysis, design specifications, and simulation results have been presented. Digital back-end processing was performed on a Xilinx RF-SoC-based ZCU1275 evaluation platform. The system used a polyphase digital architecture to enable digital beamforming of signals with up to 800~MHz of bandwidth per antenna element. 

\bibliographystyle{IEEETran}
\bibliography{comcas_new}

\end{document}